\documentclass[conference]{IEEEtran}
\IEEEoverridecommandlockouts
\usepackage{cite}
\usepackage{amsmath,amssymb,amsfonts}
\usepackage{algorithmic}
\usepackage{graphicx}
\usepackage{textcomp}
\usepackage{makecell}
\usepackage{multirow}
\usepackage{subcaption}
\usepackage{array}
\usepackage[inline]{enumitem}
\usepackage{soul}
\usepackage[table]{xcolor}

\def\BibTeX{{\rm B\kern-.05em{\sc i\kern-.025em b}\kern-.08em
    T\kern-.1667em\lower.7ex\hbox{E}\kern-.125emX}}
\begin{document}

\title{Jump off the Bandwagon? Characterizing Bandwagon Fans' Future Loyalty in Online NBA Fan Communities}

\author{\IEEEauthorblockN{Yichen Wang, Qin Lv}
\IEEEauthorblockA{Department of Computer Science, University of Colorado Boulder, Boulder, USA \\
{\{yichen.wang, qin.lv\}}@colorado.edu}
}
\maketitle

\begin{abstract}
Online user dynamics has been actively studied in recent years and bandwagon behavior is one of the most representative topics which can provide valuable insights for user identity change. Many previous studies have characterized bandwagon users and leveraged such characteristics to tackle practical problems such as community loyalty prediction. However, very few of them have investigated bandwagon dynamics from a long-term perspective. In this work, we focus on characterizing and predicting long-term bandwagon user behaviors in the context of online fan loyalty. Using a dataset collected from NBA-related discussion forums on Reddit, we trace the long-term loyalty status of bandwagon fans to capture their latent behavioral characteristics and then propose a computational model to predict their next sport season loyalty status with their home teams. Our analyses reveal that bandwagoning for most fans is a temporary switch and most of them will be back in the long term. In addition, online fans with different loyalty levels to their home teams have demonstrated different behaviors in various aspects, such as activity level, language usage and reply network properties. We then propose a model based on such behavioral characteristics to predict their next-season loyalty status. Its promising performance demonstrates the effectiveness of our behavior characterization.

\end{abstract}


\section{Introduction}

The Internet offers a variety of online content for users to explore, which brings together people with common interests to form different groups/communities. With the change of their interests and preference, users can switch between communities to participate in different activities. Loyalty is a typical term to describe this user-community relationship. In community-based platforms like Reddit, users tend to engage with more and more communities over time~\cite{tan2015all}. Characterizing these dynamics in an intra- and inter-community setting can help understand social identity change. It is also beneficial for community organizers to retain and promote community engagement~\cite{nguyen2011language,cassell2005language}.

Unlike conventional loyalty, “bandwagoning” is a special but common loyalty-related behavior in sports communities, where fans start following a sports team only because of its recent success, or certain star players in the team. This behavior brings more dynamics to users’ identities.  In the Reddit NBA community r/NBA, the bandwagon mechanism was introduced several years ago and received a lot of attention. This mechanism allows users to change their team affiliation and self-identify as bandwagon fans. During playoffs in NBA season 2016-17, 17.9\% of Cavaliers’ fans in r/NBA are bandwagon fans. Although users' intention to bandwagon within a season can be predicted~\cite{wang2020jump}, their long-term (e.g., three seasons) team affiliation status has not been studied. %Interestingly, around 80\% of the bandwagon fans are found to switch back to their original supporting team (home team) in the next season~\cite{wang2020jump}. 
Our work focuses on examining the long-term loyalty phenomenon in Reddit NBA communities. We leverage the existing structure of NBA-related discussion forums on Reddit as a testbed to study users’ bandwagon season and post-bandwagon season(s) behavior in the context of professional sports. We choose online sports fan communities because of two reasons: 1) professional sports play a significant role in modern life and a large population is actively engaged~\cite{guttmann2004ritual, Nielsen2015}; 2) professional sports teams are unambiguously competitive in nature and fans of different teams have clearly different preferences~\cite{balague2013overview,zhang2019intergroup}.

\textbf{Present work.} Our work aims to provide a thorough characterization of the bandwagon users' next-season loyalty in Reddit NBA communities. For loyalty, We categorize users into three types (loyal, partial-loyal and unloyal) based on their affiliation with their home team in the next season. Specifically, we try to answer three research questions: RQ1) What is the bandwagon fans' future-season(s) team affiliation status, i.e., does bandwagoning last long? RQ2) How do the current-season behaviors of bandwagon fans differ considering their loyalty status for the upcoming season? %What are the current-season behavioral differences of the bandwagon fans with different next-season loyalty status? 
RQ3) Can we predict their next-season loyalty status based on their current-season behavior?
Our findings suggest that bandwagoning for most users is a temporary switch, with many returning to their original team loyalties in the long term. Diving deeper into their behavior, we identify notable differences among fans of varied loyalty levels. For example, unloyal users tend to be less active but exhibit denser comment networks than loyal and partial-loyal users. Harnessing these behavioral insights, we develop a predictive model to predict bandwagon users' next-season team affiliation status.
\section{Related Work}
\textbf{Loyalty in online communities.} Loyalty in online communities has been wildly studied.% via the analysis of users' engagement. 
Users' behavioral dynamics is characterized by Tan \textit{et al.}~\cite{tan2015all}. They find that over time, users span more communities, “jump” more and concentrate less. Community evolution and the connection with its users is then studied~\cite{tan2018tracing,mensah2020characterizing}.
Hamilton \textit{et al.}~\cite{hamilton2017loyalty} further characterize loyalty by measuring user contribution and community member loyalty on both users and community level. Zhang \textit{et al.}~\cite{Zhang2017-rp} also characterize user-community interaction dynamics and show the various social phenomena manifest across communities. Human mobility and dynamics in Reddit communities are modeled in a probabilistic way by Hu \textit{et al.}~\cite{hu2019return}. As another behavior of loyalty change, users' migration is studied by Newell \textit{et al.}~\cite{newell2016user}, where motivations behind migration are identified. In addition, Zhang \textit{et al.} characterize movement and loyalty in a highly-related communities scenario~\cite{zhang2021understanding}. 
Other related behaviors like churning and retention are studied by Dror \textit{et al.} and Pudipeddi \textit{et al.}~\cite{arguello2006talk,Dror2012-yd,Pudipeddi2014-zc}, where gratitude, popularity related features, and temporal features of posts are shown to be predictive. While we learn from the aforementioned studies on online community loyalty, our work is different by focusing on a special type of loyalty, bandwagoning, and its follow-up status. This broadens the understanding of loyalty scenarios.

\textbf{Bandwagon behavior.} Bandwagon behavior is found in various fields such as politics, recommendation systems and sports. McAllister \textit{et al.}~\cite{McAllister1991-zo} apply logistic regression to survey data and find the existence of bandwagon in elections. Nadeau \textit{et al.}~\cite{nadeau1993new} find the important role of the bandwagon effect in the public opinion formation process. In recommendation systems, Sundar \textit{et al.}~\cite{sundar2008bandwagon} and Zhu \textit{et al.}~\cite{Zhu2012-zm} both find the bandwagon effect, where a user's opinion is swayed by other users' choices.  Bandwagon theory in professional sports is used to explain fans' movement~\cite{wann1990hard}. Wang \textit{et al.}'s work~\cite{wang2020jump} on characterizing NBA fan bandwagon behavior is the first quantitative study on the online bandwagon fan behavior, and they build a model to predict if a fan will bandwagon during playoffs based on pre-playoffs behavior. While previous studies on bandwagon behavior provide valuable insights, most are primarily confined to short-term behaviors. Setting our study apart, we extend the timeline by focusing on the long-term loyalty trends of 'already-bandwagon' users in future seasons. Subsequently, drawing from the insights on long-term behavior our study unveiled, we have developed a computational model to predict the next-season team affiliations of bandwagon users.

\textbf{Sports fan behavior.} Online sports communities are good testbeds to study online community loyalty \cite{wang2020jump, zhang2018we}. As a crucial part of sports team management, sports fan behavior has received a lot of attention. Fan loyalty is studied by Yoshda \textit{et al.} and Bee \textit{et al.} using model-based approaches~\cite{bee2010exploring,yoshida2015predicting}. Fan attraction, psychological commitment, and resistance to change are found to be useful predictors for loyalty~\cite{bee2010exploring}. The relationship between team loyalty, sponsorship awareness, attitude toward the sponsor, and purchase intentions is studied~\cite{biscaia2013sport}. Beyond team management implications, sports fan behavior can also be used as a lens to study humans. The gender difference when considering being a sports fan is studied by Beth \textit{et al.}~\cite{dietz2000sex}. NFL fans' Twitter data is used to study group conversations~\cite{margolin2015conversing}. Building upon these studies and using online sports communities as testbeds, our research specifically navigates the online NBA community's fan behavior to better understand the bandwagon phenomenon.

\section{Dataset Construction}
\subsection{Collecting NBA-related Subreddits Data}
We focus on NBA-related discussion forums on Reddit, an active community-driven platform where users can participate in different communities to submit posts and comments. On Reddit, discussion forums are called Subreddits and denoted by ``r/'' followed by the community's name, and r/NBA is the main forum for users to discuss anything related to the league. There are 30 teams in the league and each team has its own subreddit, e.g., r/lakers. We focus on 4 NBA seasons: 2015-16, 2016-17, 2017-18, and 2018-19. Seasons before 2015 are not considered because the bandwagon mechanism has not been introduced then. The most recent seasons (season 2019-20, 2020-21, 2021-2022) are not considered because they were highly impacted by the COVID-19 pandemic: the 2019-2020 season was suspended for several months, and the rules for playoffs team eligibility has been changed since then, which could significantly change the normal discussion pattern in the related subreddits. We obtained all the 0.66M posts and 41M comments during the four seasons in the 31 NBA-related subreddits from Pushshift~\cite{baumgartner2020pushshift} after filtering out posts and comments authored by deleted users (replies towards deleted users were not filtered). %There are 0.66M posts and 41M comments in total across the 4 seasons in our dataset. 

\subsection{Identifying Bandwagon Fans}
\begin{figure*}[htbp]
\centering
%\vspace{-0.5em} 
    \begin{subfigure}{0.5\linewidth}
    \centering
      \includegraphics[width=0.91\textwidth]{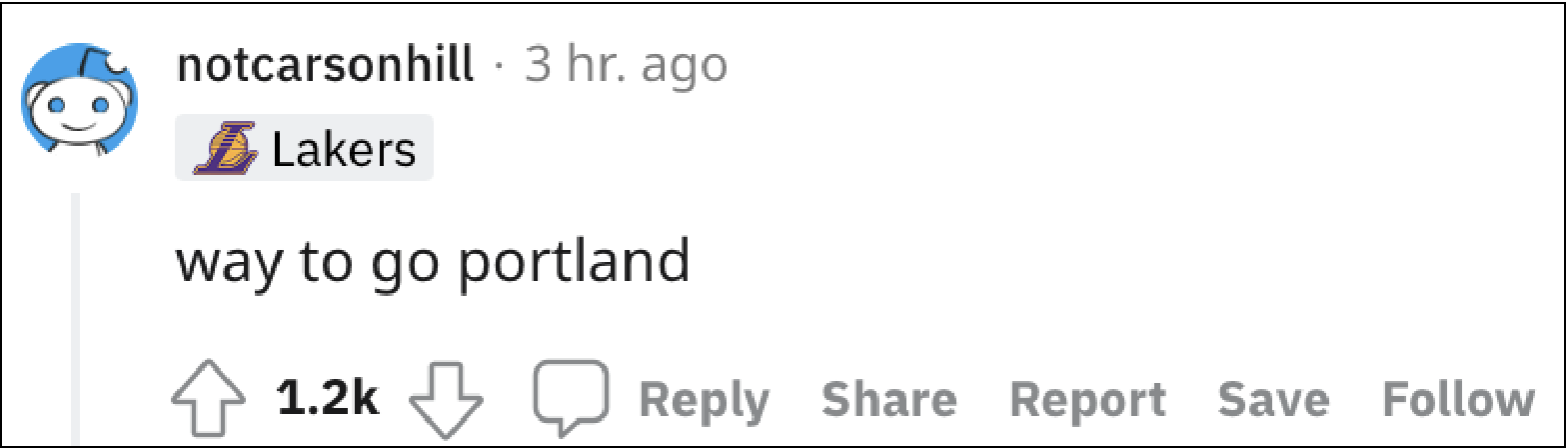}
      \caption{An example of (normal) flair}
      \label{fig:flair}
    \end{subfigure}\hfill
    \begin{subfigure}{0.5\linewidth}
    \centering
      \includegraphics[width=0.84\textwidth]{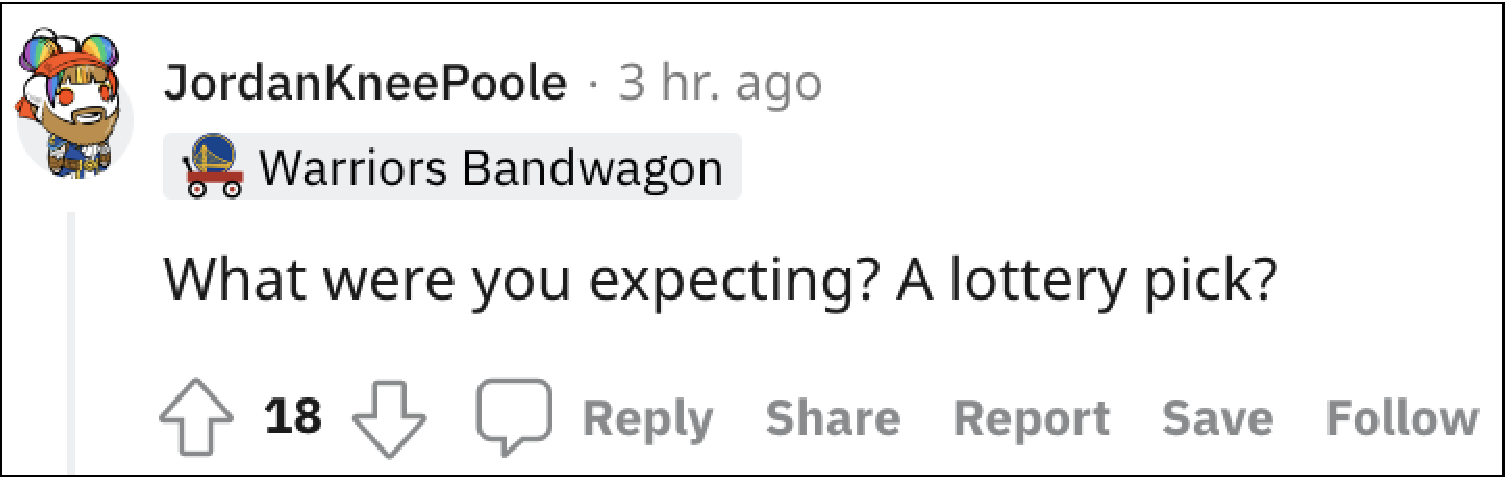}
      \caption{An example of bandwagon flair}
      \label{fig:bw_flair}
    \end{subfigure}
%\captionsetup{justification=centering,margin=2cm}
\caption{Examples of flair usage in r/NBA}
\label{fig:flair_example}
\end{figure*}

In r/NBA, a user's team affiliation can be acquired directly through a mechanism known as "flair". Flair appears as a short text next to the username in posts and comments (Fig.~\ref{fig:flair}). In r/NBA, users can choose to use flairs to indicate support and each user cannot have multiple flairs at the same time. Flair can be changed as many times as one wants. During playoffs, some users change their flairs to a bandwagon flair of a different team (Fig.~\ref{fig:bw_flair}). This self-chosen (bandwagon) flair is a useful sign reflecting a user's team affiliation status. Posting/commenting in r/NBA with a team's flair is considered as supporting that team.

%Making use of (non-bandwagon) flairs, we define \textit{fan} of a team. Then we use users' bandwagon flairs to identify bandwagon fans
%---------------------------------------------------------------------
%---------------------------------------------------------------------

Same as the method in~\cite{wang2020jump}, we define a user as a \textit{fan} of a team during a specific period of time if the user indicates support (flair) only for that team and such support sustains over all activities (posts/comments) during that time period. Note that this flair does not contain the word "bandwagon". A \textit{bandwagon fan} means a user is a \textit{fan} of a team $T_A$ until a time point, and then switches the flair to team $T_B$'s bandwagon flair, %($T_B \neq T_A$, so bandwagon flair $T_A$ is regarded as the same as flair $T_A$), 
regardless of any other bandwagon flair changes thereafter. There are 2565, 1461 and 995 \textit{bandwagon fans} in season 2015-16, 2016-17 and 2017-18, respectively, and they are the target users of this study. \footnote{The future behaviors of 2018-19 \textit{Bandwagon fans} are not studied since as the future season of 2018-2019, season 2019-20 is disrupted by COVID-19}.%this work focuses on bandwagon fans' future-season loyalty status, while the data in season 2019-20 is disrupted by COVID-19.} %As season 2019-20 is disrupted by COVID-19 , the flair usage in this season may be very different and can be studied as a future work.}

\subsection{Defining Bandwagon Fans' Loyalty in Future Season(s)}
We define home team of a bandwagon fan as the team they originally support (having the team flair) before bandwagoning. %To answer this research questiacon, we first examine all the bandwagon fans' flairs in the \textbf{next} season after their bandwagon season. 
Based on how long they support their home team in the next season, we categorized the bandwagon fans into three types: 
\begin{itemize}
    \item \textit{Loyal} bandwagon fans change their flairs back to the home team in the next season and sustain throughout that season. For example, if a user was team A's fan and team B's bandwagon fan in season 2015-16, and goes back to be team A's fan in the whole season of 2016-17, then this user is categorized as a \textit{loyal} bandwagon fan.
    \item \textit{Partial-loyal} bandwagon fans change their flairs back to home team for some time in the next season and then leave again.
    \item \textit{Unloyal} bandwagon fans never return to their home team (they could stay with the bandwagon team, or change to another team). 
\end{itemize}
There are also a number of fans completely leaving r/NBA and we do not consider these users, as their team affiliation is unknown due to no activity. The number of different types of bandwagon fans is shown in Table~\ref{tab:count_fan_type}.
%---------------------------------------------------------------------
%\begin{table}[h]
\begin{table}%{r}{0.45\textwidth}
\centering
\caption{Number of different bandwagon fans}
\label{tab:count_fan_type}
\begin{tabular}{|c|c|c|c|}
\hline
Season &  2015-16 &  2016-17 &  2017-18 \\%\backslashbox{Type}{Season}
\hline
    Loyal & 1894  & 878 & 822 \\
    Partial-loyal & 417  & 356 & 122 \\
    Unloyal & 254 & 227 & 51\\
\hline
    Total & 2565 & 1461  & 995\\
\hline
\end{tabular}
\end{table}
%---------------------------------------------------------------------
\section{RQ1: What is the bandwagon fans' future-season(s) team affiliation status?}
\begin{figure*}[htbp]
    %\vspace{-0.7em}
    \begin{subfigure}{0.5\linewidth}
    \centering
      \includegraphics[width=0.7\textwidth]{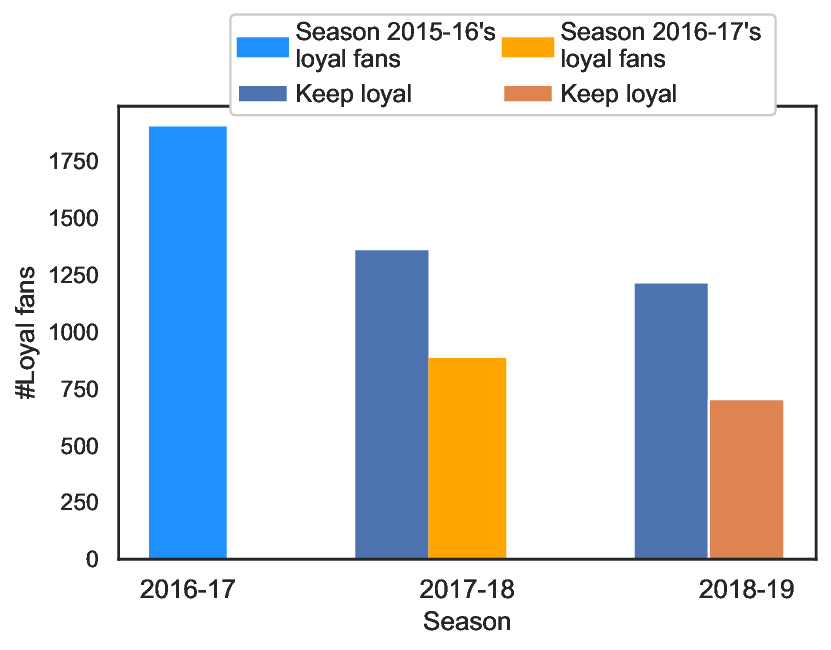}
      \caption{Loyal bandwagon fans' long-term retention}
    \end{subfigure}\hfill
    \begin{subfigure}{0.5\linewidth}
    \centering
      \includegraphics[width=0.7\textwidth]{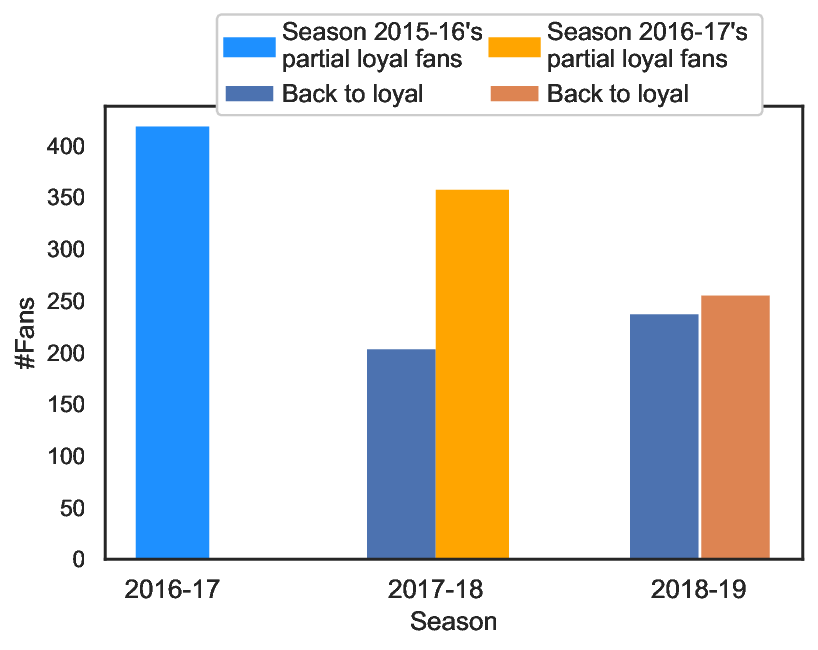}
      \caption{Partial-loyal bandwagon fans' long-term retention}
    \end{subfigure}
%\captionsetup{justification=centering,margin=2cm}
\caption{Bandwagon fans' longer term retention rate.}
\label{fig:rq1_retention}
\end{figure*}

\textbf{Majority of the bandwagon fans are still loyal in the next season(s).} We find that a large portion of the bandwagon fans are still loyal in the future seasons, which indicates that bandwagoning is a temporary diversion for most bandwagon fans, and it doesn't mean the loss of interest in the home team. Having the next-season loyalty status, we looked into bandwagon fans' longer term retention of their home team. For bandwagon fans identified in season 2015-16, we were able to check their flairs in the next three seasons. For bandwagon fans in season 2016-17 we checked their loyalty in the next two seasons. For season 2015-16's loyal bandwagon fans, their retention rates are 71.9\% and 64.1\% in season 2017-18 and 2018-19, respectively. For season 2016-17's bandwagon fans who are still loyal in season 2017-18, their retention rate is 80.0\% in season 2018-19. This pattern also applies to partial-loyal bandwagon fans. 48.9\% and 57.1\% of 2015-16's partial-loyal fans' were back to loyal in seasons 2017-18 and 2018-19, respectively. Fig.~\ref{fig:rq1_retention} shows their long term retention status. These results suggest that r/NBA bandwagon fans should not be treated as vagrant users as their leaving is not permanent and their loyalty is quite strong over time.

%---------------------------------------------------------------------
%---------------------------------------------------------------------

\begin{figure*}[t]
\centering
    \begin{subfigure}{0.326\linewidth}
      \includegraphics[width=\textwidth]{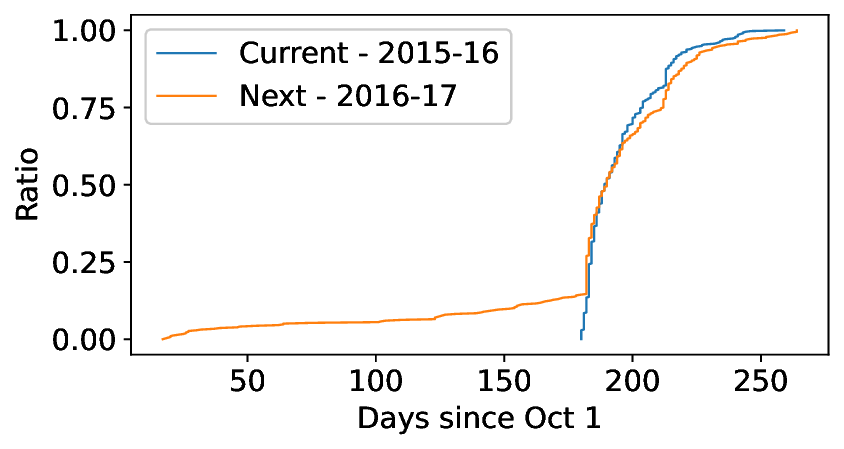}
      \caption{Season 2015-16 partial-loyal fans' flair change time}
      \label{fig:rq1_time1}
    \end{subfigure}\hfill
    \begin{subfigure}{0.326\linewidth}
      \includegraphics[width=\textwidth]{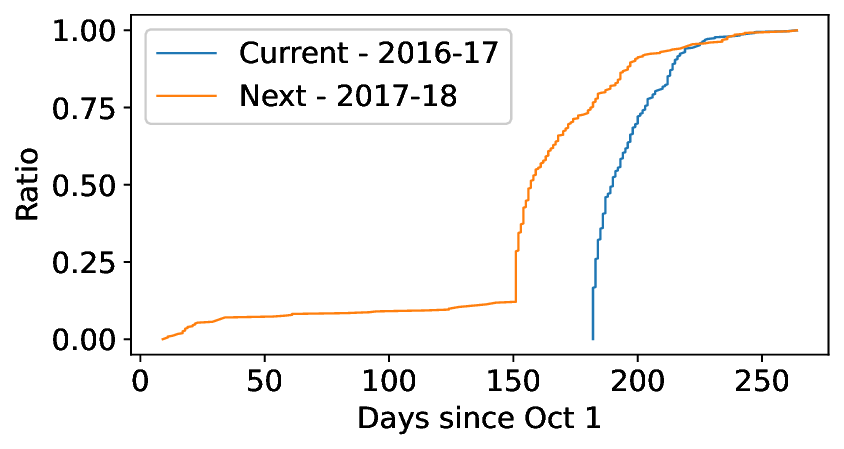}
      \caption{Season 2016-17 partial-loyal fans' flair change time}
      \label{fig:rq1_time2}
    \end{subfigure}
    \begin{subfigure}{0.326\linewidth}
      \includegraphics[width=\textwidth]{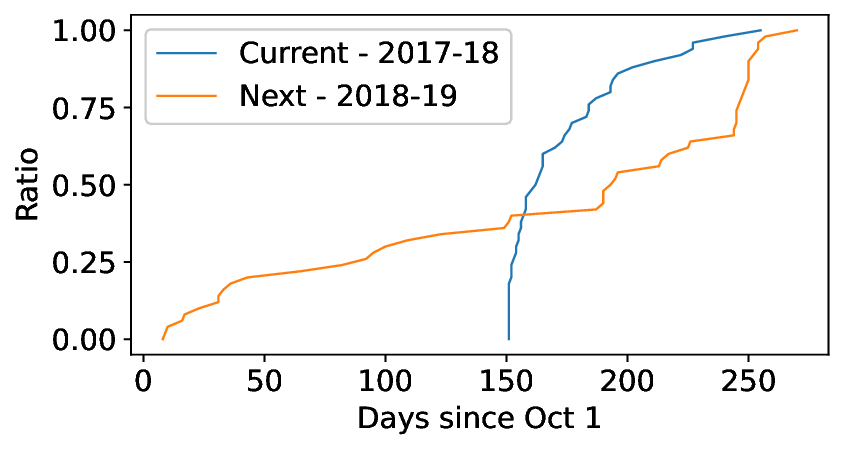}
      \caption{Season 2017-18 partial-loyal fans' flair change time}
      \label{fig:rq1_time3}
    \end{subfigure}
\caption{Partial-loyal bandwagon fans' flair change time}
\label{fig:rq1_time}
\end{figure*}

\textbf{Partial-loyal fans come from weaker teams.} It is found that better team attract more bandwagon fans, but the relationship with home team performance is unclear~\cite{wang2020jump}. To further explore this direction, we looked into the performance of home teams within the bandwagon fan. We compare the home team standings across different bandwagon fan categories via t-tests, and the partial-loyal users' home team is significantly weaker than others (9.1 vs* 9.5 vs* 8.7\footnote{Throughout this paper, the number of stars indicate p-values: $^{***}:p < 0.001$, $^{**}:p < 0.01$, $^{*}:p < 0.05$. All the significant tests are after Bonferroni correction.}). Note that a large team standing means a weaker team performance. This result reveals that more ``swing'' bandwagon fans come from weaker teams.
%\vspace{-1em}
\textbf{Bandwagon flair switch time is getting earlier in future season(s).} Another finding about bandwagon fans is the time to switch to a bandwagon flair. We examined each partial-loyal bandwagon fan's first activity (post/comment) time after changing the home team flair in their current and next season. Then we calculate the days between the first non-home team flair activity and Oct 1. Oct is the month when the NBA season starts. As each season starts on a different day, we use Oct 1 to align the time differences. Fig.~\ref{fig:rq1_time} shows the CDF plot of the current and next seasons' first non-home team flair activity day of season 2015-16's partial-loyal bandwagon fans. A number of them switched to bandwagon flair very early in the next season, which means that some fans lost interest in their home team at the very early stage of the next season. This phenomenon is more common since season 2017-18 (Fig.~\ref{fig:rq1_time2},\ref{fig:rq1_time3}), when the fans change their flair earlier.

\section{RQ2: How do the current-season behaviors of bandwagon fans differ considering their loyalty status for the upcoming season?}
To understand the signs of being loyal or not in the next season, we first characterize how the three types of bandwagon fans behave differently in the current season. Specifically, behaviors of three aspects are studied: activity (post/comment, the feedback the receive), language usage, and network properties. T-test was also used to compare the behavioral features. Note that we use ``fans'' to represent the bandwagon fans we study in the following sections, omitting the word ``bandwagon'' for conciseness.
%---------------------------------------------------------------------
%---------------------------------------------------------------------

\subsection{Activity}

\begin{figure*}[h]
\centering
%\vspace{-2em} 
    \begin{subfigure}{0.328\linewidth}
      \includegraphics[width=0.9\textwidth]{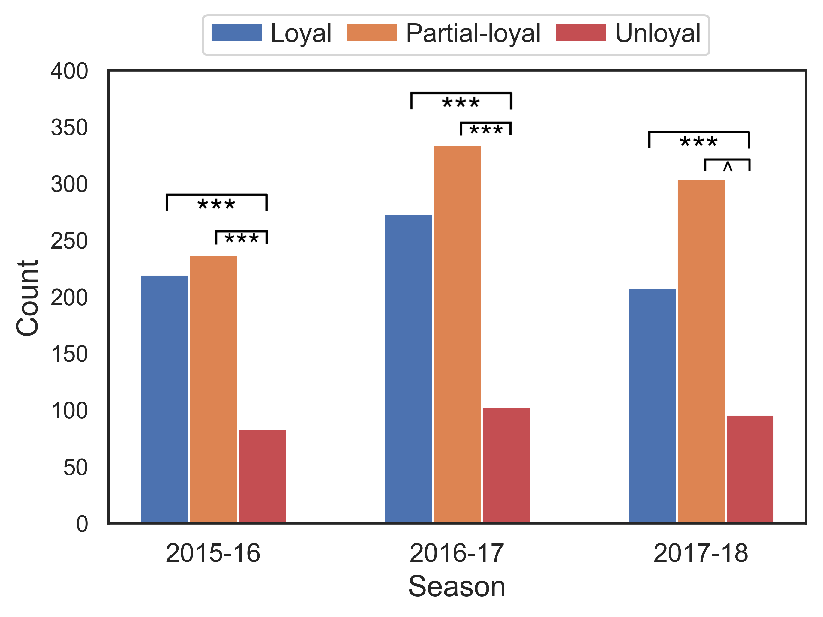}
      \caption{Average activity count in r/NBA \\ \quad\quad}
      \label{fig:rq2_count_activity}
    \end{subfigure}\hfill
    \begin{subfigure}{0.328\linewidth}
      \includegraphics[width=0.9\textwidth]{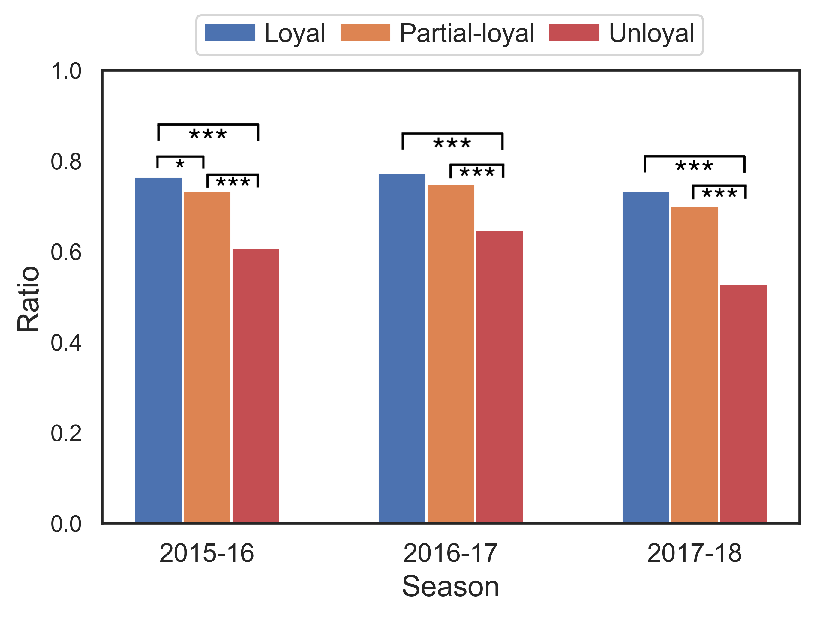}
      \caption{Average ratio of comments posted with home team flair in r/NBA}
      \label{fig:rq2_ratio_comments_hometeam}
    \end{subfigure}\hfill
    \begin{subfigure}{0.328\linewidth}
      \includegraphics[width=0.9\textwidth]{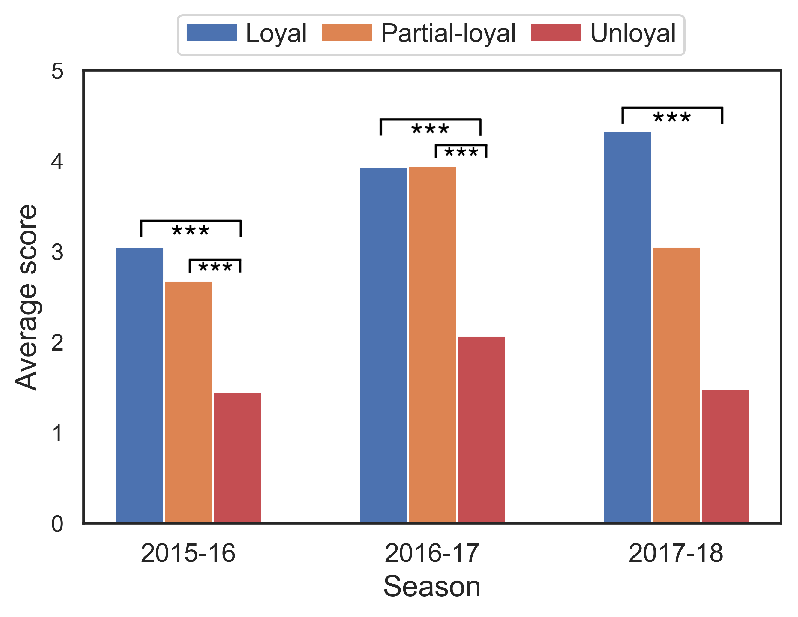}
      \caption{Average score of comments in home team subreddit}
      \label{fig:rq2_avg_score}
    \end{subfigure}
%\captionsetup{justification=centering,margin=2cm}
\caption{Activity related feature comparison of different bandwagon fans in three seasons.\protect\footnotemark}
\label{fig:rq2_activity}
\end{figure*}

\textbf{Loyal and partial-loyal fans are more active than unloyal fans, and devote more time to their home teams.} We compared their activity related features in both r/NBA and team subreddits under many scenarios. First, we found loyal and partial-loyal fans have more activities in both r/NBA and team subreddits (e.g., Fig.~\ref{fig:rq2_count_activity} is the result in r/NBA).  To measure their attachment to their home teams, we further examined the ratio of comments with home team flair in r/NBA, and the ratio of comments in the home team subreddit, and the results (e.g., Fig.~\ref{fig:rq2_ratio_comments_hometeam} is the result in r/NBA) are consistent with their next-season loyalty level, where more loyal fans have longer support for their home teams.

\textbf{Loyal and partial-loyal fans receive better feedback in their home team subreddits.} Here, we compute a score for each comment (\#upvotes - \#downvotes) as a measure of received feedback. Using this metric we looked into the feedbacks they received in home and bandwagon team subreddits. We find that the scores of loyal and partial-loyal fans are higher than that of unloyal fans in their home team subreddits (Fig.~\ref{fig:rq2_avg_score}), which suggests that these more-loyal fans are more welcomed in their base community.

%---------------------------------------------------------------------

\footnotetext{Throughout this paper, the significance of the t-test of the behavioral features between different fans are shown as the connection above the bars and p-values are indicated by the symbols: $***:p < 0.001$, $**:p < 0.01$, $*:p < 0.05$, $\wedge:p < 0.1$. All t the results are after Bonferroni correction.}
%---------------------------------------------------------------------
%\vspace{-2em} 
\subsection{Language Usage}
Although we use both posting and commenting as indicators of user activities, when studying language usage we focus on comments since posts are more about news and game reports, and less reflective of users' personal opinions.

\textbf{There is no significant difference in the lexicon usage, except for \textit{we} pronouns.}
To capture user language’s lexical style, we calculated the proportion or summary variable of different types of words in their comments using LIWC word categories~\cite{boyd2022development}, a well-known set of word categories that were created to capture people’s social and psychological states. To our surprise, the only lexicon that differs is \textit{we} pronouns, where the loyal and partial-loyal fans used more than unloyal fans, %(Fig.~\ref{fig:rq2_we} in appendix), 
indicating that the more loyal fans'language shows more sense of belongings. Different from previous work on intergroup fans being more rude and negative~\cite{zhang2019intergroup}, this result shows that under bandwagon scenario these fans with different home team affiliation levels have similar language usage.

\textbf{Loyal fans focus more on the technical side while the other two groups are more emotional.}
We identify the distinguishing words that are overused by different bandwagon fans. To achieve this, we apply the fightin-words algorithm with the informative Dirichlet prior model~\cite{monroe2008fightin}. This model is known to outperform other traditional methods in detecting word usage differences between corpora, such as Pointwise Mutual Information~\cite{manning1999foundations} and TF-IDF~\cite{chowdhury2010introduction}, by not over-emphasizing fluctuations of rare words. Table~\ref{tab:fightin-words} shows the results for season 2015-16. Our results show that loyal fans focus more on the sport itself and use more technical words such as ``defense'', ``perimeter''. In comparison, the other two groups are more emotional and use more swear words. We also conducted language analysis using lexicon based method, but did not find significant difference. As most swear words / slangs are used in acronyms, they are hard to be captured by the lexicons, thus not giving a very significant result in the lexicon based analysis.

\begin{table*}[t]
\centering
\caption{Top 10 overused words by different bandwagon fans in season 2015-16 (Words of specific player/team names have been removed, so that we can focus on users' own personalities. All the words have a z-score larger than 5.)}
\label{tab:fightin-words}
\begin{tabular}{|c|c|}
\hline
Fans type & Overused words (Words inside the parentheses are explanations)\\
\hline
 Loyal  &\makecell{postseason, f5 (to look at something obsessively and repeatedly), elite, \\ bc (because), perimeter, defense, offensive, value, offense, defender} \\
\hline
Partial-loyal &\makecell{god, oh, upvote, gonna, lmfao, \\ffs (for fuck's sake), specially, photo, photographer, hoist} \\
\hline                                
Unloyal & \makecell{wow, wtf, u/user\_simulator (a summonable Reddit bot), percentage,\\   lastly, mofo (motherfucker), aussie, school, imo, nahh}\\
\hline
\end{tabular}
\end{table*}

\subsection{Network properties}
Finally, we examine the properties of bandwagon fans’ ego networks in r/NBA by constructing a directed replying network between users in r/NBA. Specifically, each node in the network represents a user, and there is an edge from node A to node B if user A replied to user B, and vice versa. For each bandwagon fan, we build the network edges based on the neighbors, i.e., whom they replied to and got replies from, and the replies between the neighbors.

%---------------------------------------------------------------------
\begin{figure*}[h]
%\centering
%\vspace{-1em} 
    \begin{subfigure}{0.325\linewidth}
    \includegraphics[width=0.9\linewidth]{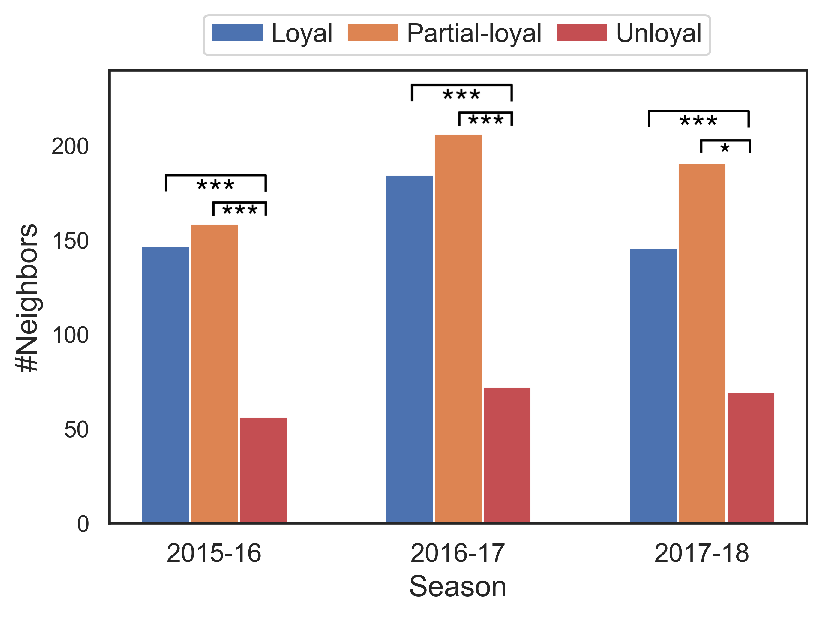}
    \caption{Average number of neighbors in the ego-network}
    \label{fig:rq2_neighbor}
    \end{subfigure}\hfill
    \begin{subfigure}{0.325\linewidth}
    \centering
      \includegraphics[width=0.9\textwidth]{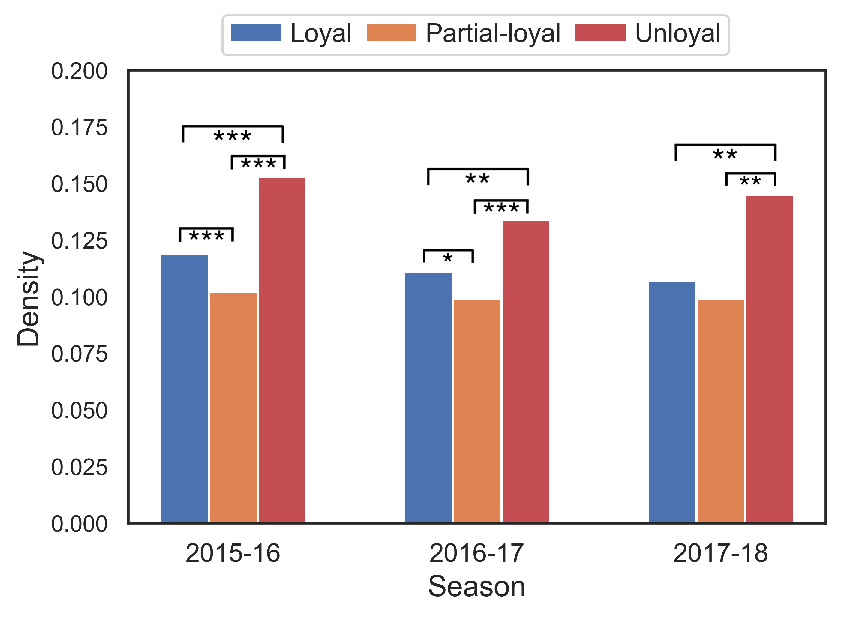}
      \caption{Average density of the ego-network\\ \quad}
      \label{fig:rq2_density}
    \end{subfigure}\hfill
    \begin{subfigure}{0.325\linewidth}
    \centering
      \includegraphics[width=0.9\textwidth]{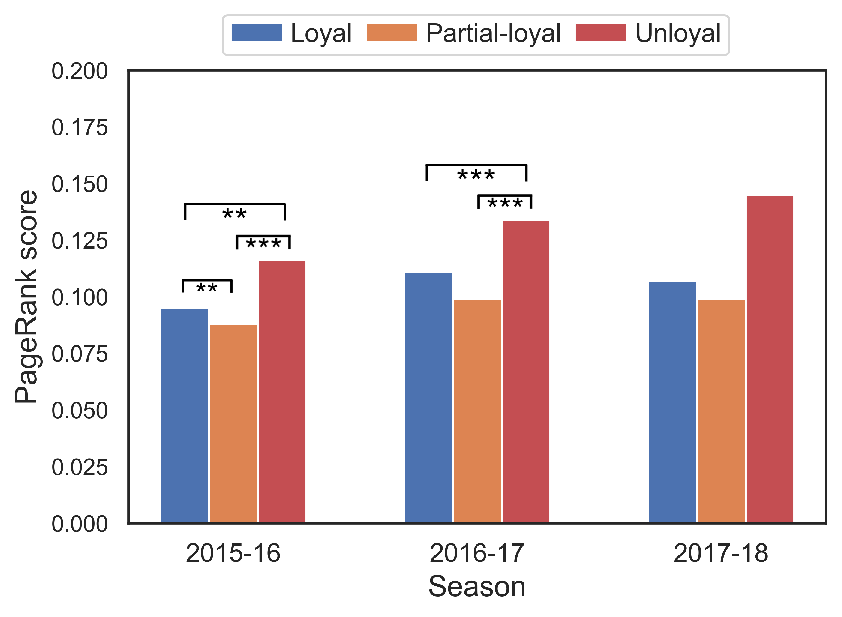}
      \caption{Average PageRank score of each bandwagon fan in their ego-network}
      \label{fig:rq2_pagerank}
    \end{subfigure}
%\captionsetup{justification=centering,margin=2cm}
\caption{Network related feature comparison of the different bandwagon fans in the 3 seasons.}
\label{fig:rq2_network}
\end{figure*}
%---------------------------------------------------------------------
%\vspace{-3em}
\textbf{Loyal and partial-loyal fans have more neighbors in their ego-networks, but their ego-networks are less clustered.} We first find that loyal and partial-loyal fans have more neighbors than unloyal fans (Fig.~\ref{fig:rq2_neighbor} in appendix), i.e., higher in and out degrees in their replying networks. As they have higher activity level, it is natural that they communicate with more people. However, we find that unloyal user networks have higher network density and PangRank score~\cite{page1999pagerank}, as shown in Fig.~\ref{fig:rq2_network}. These observations suggest that although loyal and partial-loyal fans connect with more users, the connection between their neighbors are sparser, compared with the neighbors of unloyal fans, which means unloyal fans are communicating with users who are more tightly connected. In addition, a higher PageRank score demonstrates that unloyal fans have higher possibility to be visited when tracing the reply trajectory, indicating that they communicate with more influential users in the r/NBA community than loyal and partial-loyal fans. We also find that unloyal fans have less communication with users from the same team. This finding differs from previous work on the existence of dense clusters within partisan groups in political blogs and Twitter~\cite{adamic2005political,ausserhofer2013national}. A possible explanation is that these users are bandwagon fans, who are already a group of less loyal fans compared with those who do not choose to bandwagon, thus their communication network is not a partisan group. Another finding is that some of the differences are not significant in season 2018-19, such as average score of comments in home team subreddit, PageRank score and \textit{we} pronoun, which implies that users' behavioral pattern changes a lot in that season. %This is also reflected in their flair change time as discussed in Section 4.

\section{RQ3: Can we predict users' next-season loyalty status based on their current-season behavior?}
We consider the prediction task of determining which home team affiliation category (loyal, partial-loyal and unloyal) a bandwagon fan will fall into in the next season. The goal of this prediction task is to demonstrate that future loyalty status can be inferred from a fan’s current-season behavioral features observed. We use the features of the bandwagon fans in the season they choose to bandwagon (current season) to predict their next-season home team affiliation category. For example, for the bandwagon fans in season 2015-16, their behavioral features in season 2015-16 are the model input and their home team affiliation in season 2016-17 is the output. We use the data of the bandwagon fans in season 2015-16 and 2016-17 as the training set and the data of 2017-18 bandwagon fans as the test set. The following feature groups are experimented:
\begin{itemize}
    \item Activity: the activity count and the ratio of comments with home team flair in r/NBA; activity count in (all) the team subreddits; the ratio of comments and the average score of comments in the home team subreddit
    \item Network: number of neighbors; density and PageRank score.
    \item Comment (baseline): the TF-IDF features of each user's comments in r/NBA. 
    \item Subreddit: the bag of subreddits (excluding NBA related ones) visited. It is inspired by the work on user movement in highly related communities~\cite{zhang2021understanding}.
\end{itemize}

\begin{figure*}[h]
    \begin{subfigure}{0.33\linewidth}
      \includegraphics[width=\textwidth]{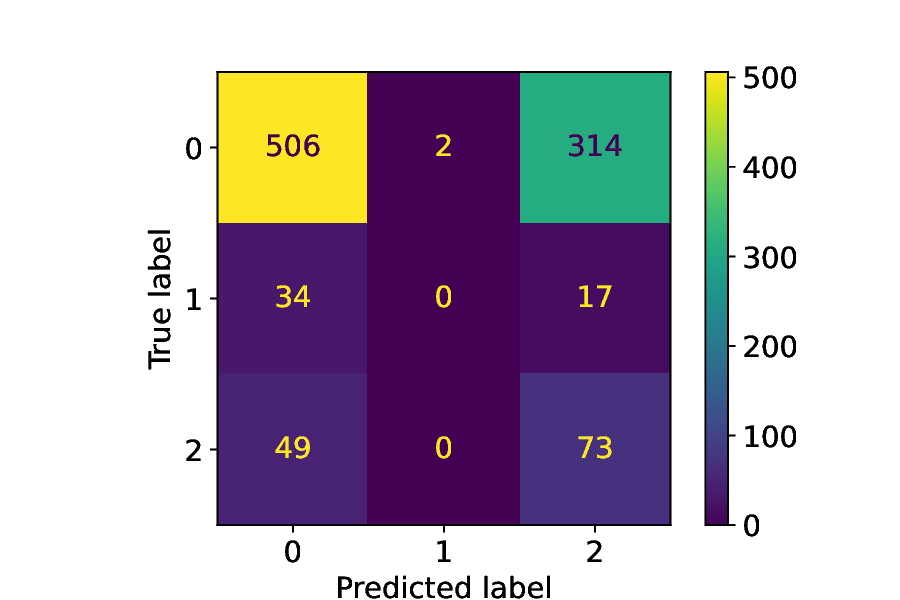}
      \centering
      \caption{Comment}
    \end{subfigure}\hfill
    \begin{subfigure}{0.33\linewidth}
    \centering
      \includegraphics[width=\textwidth]{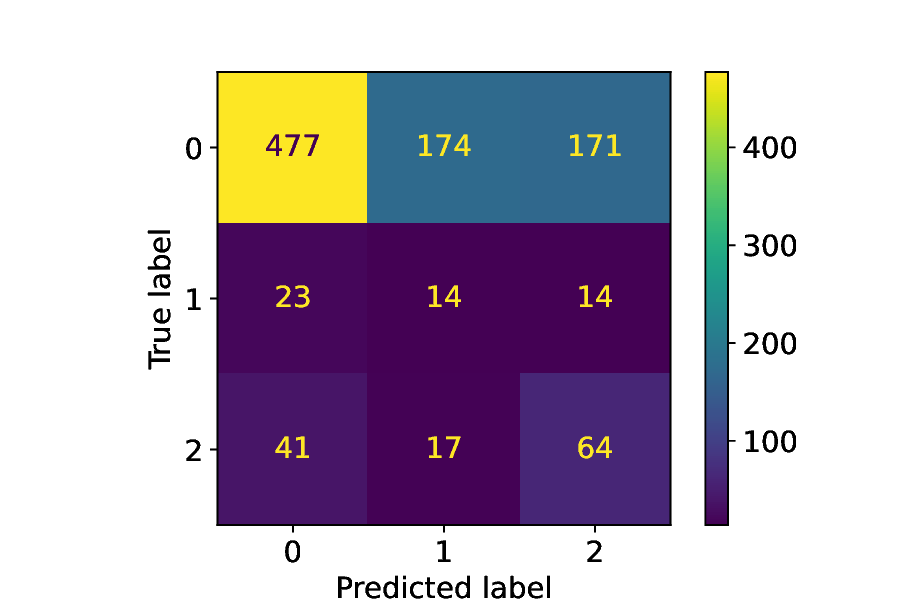}
      \caption{Activity \& network}
    \end{subfigure}\hfill
    \begin{subfigure}{0.33\linewidth}
    \centering
      \includegraphics[width=\textwidth]{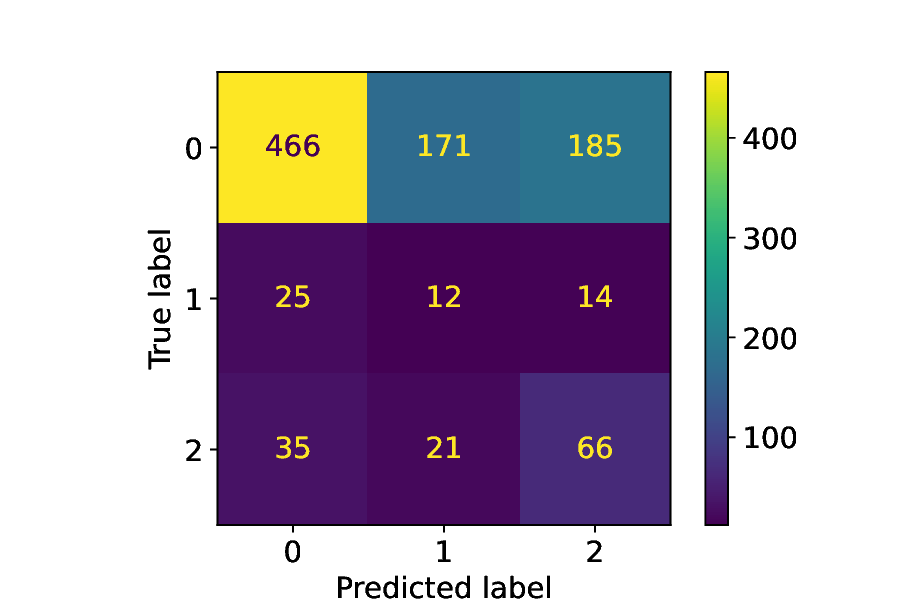}
      \caption{Activity only }
    \end{subfigure}
    \begin{subfigure}{0.33\linewidth}
    \centering
      \includegraphics[width=\textwidth]{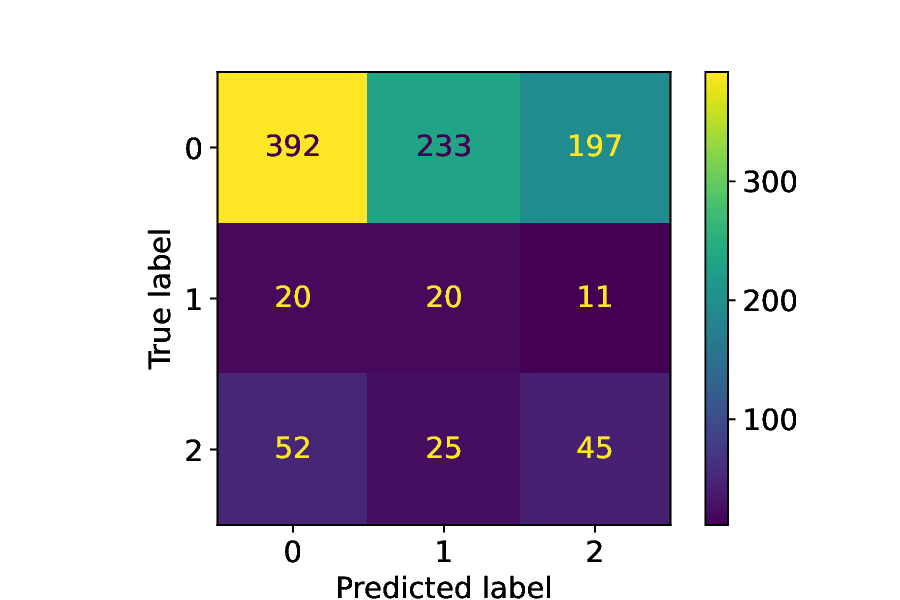}
      \caption{Network only}
    \end{subfigure}\hfill
    \begin{subfigure}{0.33\linewidth}
    \centering
      \includegraphics[width=\textwidth]{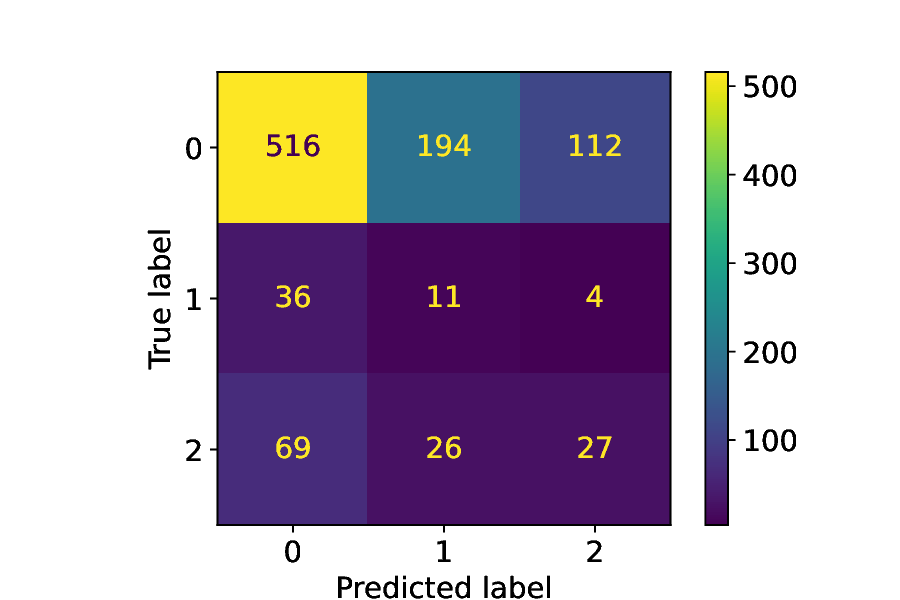}
      \caption{Bag-of-subreddit}
    \end{subfigure}\hfill
    \begin{subfigure}{0.33\linewidth}
    \centering
      \includegraphics[width=\textwidth]{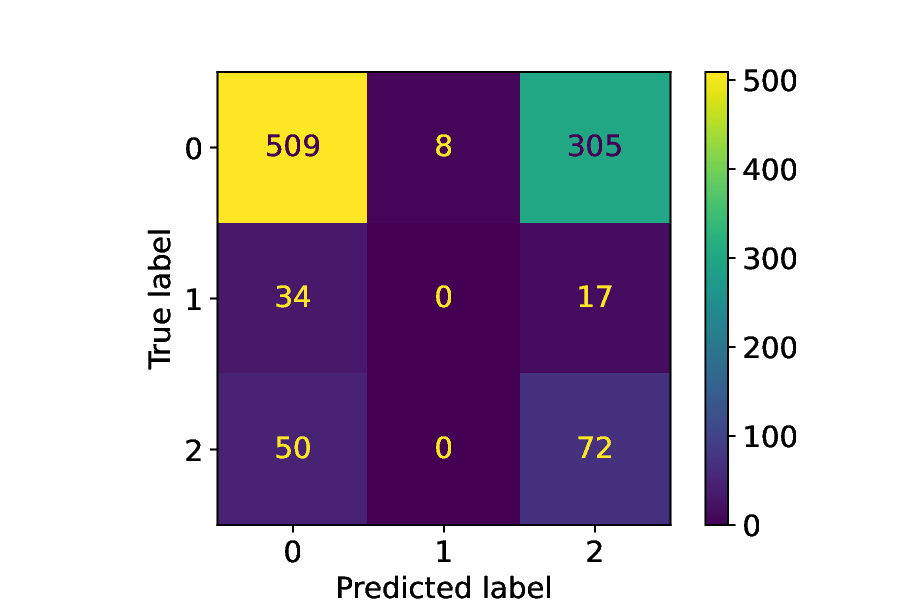}
      \caption{All features}
    \end{subfigure}\hfill
\caption{Confusion matrices of results on 3-class prediction task with different feature groups. 0 - Loyal, 1 - Partial loyal, 2 - Unloyal.}
\label{fig:rq3_multiclass}
\end{figure*}

\begin{figure}[h]%{r}{0.45\textwidth}
% \vspace{-2em}
\includegraphics[width=0.45\textwidth]{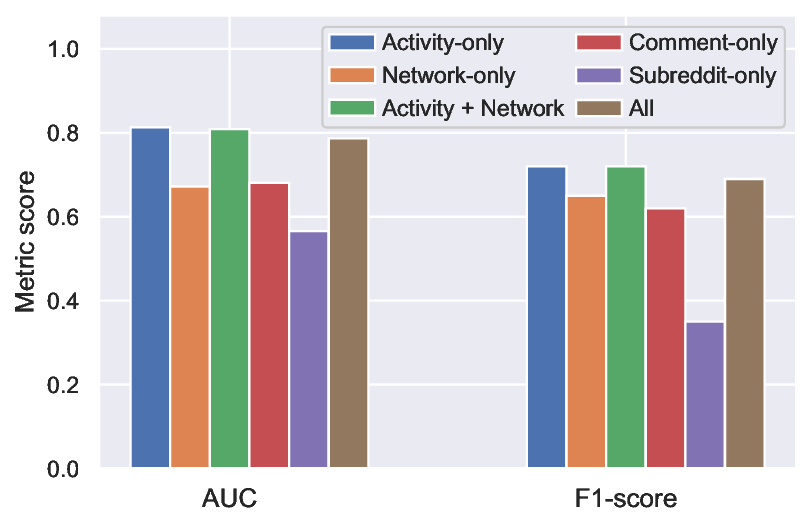}
\caption{Model performance on the binary classification task.}
\label{fig:rq3_binary}
\end{figure}

A random forest model was used to conduct the training and prediction. Random search was used to find optimal model parameters. We first did experiments on the 3-class prediction task. Fig.~\ref{fig:rq3_multiclass} shows the confusion matrix of the results. The performance on partial-loyal fans is not good, The reason is that many partial-loyal fans are quite ``loyal'', proved by their high retention rate in long term (Fig.~\ref{fig:rq1_retention}). Therefore, we separated the partial-loyal fans into two parts based on the ratio of comments with home team flair. Fans with ratio more than 50\% were merged into loyal fans and others were merged into unloyal fans. Then we conducted experiments on a binary classification task. We also randomly downsampled the loyal fans to the same number of unloyal fans in the test set to make a balanced evaluation. Fig.~\ref{fig:rq3_binary} summarizes the results. First, the activity and network features can effectively predict bandwagon fans' next-season loyalty status, which greatly improves the baseline AUC by 16\% and F1-score by 19\%. Second, activity features have more prediction power than network features, which aligns with the non-significant observation of the PageRank score in season 2017-18 (Fig.~\ref{fig:rq2_pagerank}).

%---------------------------------------------------------------------
%\begin{table}[t]
%\centering
%\caption{Model prediction performance comparison}
%\label{tab:results_binary}
%\begin{tabular}{|c|c|c|c|c|c|c|}
%\hline
%\backslashbox{Metric}{Features} & Activity-only & Network-only & Activity + Network & Comment-only & Subreddit-only & All\\
%\hline
%AUC & \textbf{0.81}  & 0.67 & \textbf{0.81} & 0.68 & 0.57 & 0.1 \\
%\hline
%F1 & \textbf{0.72}  & 0.65 & \textbf{0.72} & 0.62 & 0.35 & 0.1 \\
%\hline
%\end{tabular}
%\end{table}
%---------------------------------------------------------------------

\section{Concluding Discussion}
In this paper, we focus on bandwagon fans' future-season(s) loyalty status. We categorize the bandwagon fans into 3 types based on their home team affiliation in the future season(s). We first find that majority of the bandwagon fans are still loyal in the future seasons and only a small portion of them completely leave their home team. We then find that loyal and partial loyal fans are more active than unloyal fans, and loyal fans are less emotional and care more about the sports itself. Unloyal fans are also found to have denser reply network and higher importance. Making use of the observed features, we build a model which can effectively predict bandwagon fans’ next-season loyalty status.  As far as we know, this is the first work focusing on bandwagon fans' future loyalty.

\textbf{Implications.} Our analysis shows that bandwagon behavior is temporary and fans' longer-term loyalty is predictable via their activity characteristics, which demonstrates the power of behavior analysis to understand users. The results of the first research question indicate that temporary bandwagoning in NBA communities should not be regarded as a pure unfaithful behavior, suggesting that the ``jump more, concentrate less'' pattern~\cite{tan2015all} may not apply in sports communities. %like the conventional loyalty problem,  
This could encourage more studies in the short-term community engagement. As the bandwagon status is derived from users' self-chosen team flairs, the finding raises the questions in social identity theory~\cite{deaux1993reconstructing}: Under what scenarios the explicit self-identification can represent a user's implicit, real opinion, and under what scenarios is the identification not accurate?

In addition, the findings in the network features show that unloyal fans are more ``important'' in their ego-network (connected with more influential users), although they are less active than loyal fans. This brings a question that what type of user is more influential in an inter-group community. Our model also demonstrates the possibility of using behavioral features to predict bandwagon fans' future loyalty status. It is crucial for community managers to identify potential loyal and vagrant fans with the goals of maintaining their fan base. Our results show the potential of applying model based approach to automate this work.

Moreover, as bandwagoning is found in other fields such as political election~\cite{McAllister1991-zo}, and people could have similar preference change trace as in sports, our work could potentially be applied in these areas to characterize the people who are easily swayed by others' opinions.

Last but not least, it is found that the bandwagon flair is deviating from its original goal in the long term. For instance, bandwagon fans are decreasing over time, and the bandwagon flair usage is becoming more and more arbitrary. The bandwagon flair is even used by some users during the very beginning of a season, which is not the original intention for encouraging engagement during playoffs. Although prior work has shown that incorporating group identity can help strengthen member attachment in online communities~\cite{ren2012building}, this bandwagon flair mechanism does not seem to work very well in the long term. This brings up an important challenge for community managers to tackle: how to devise more effective mechanisms to enhance user participation.

\textbf{Limitations and future work.} The first limitation of our work is how representative our dataset is. Although r/NBA is now playing an important role among fans, the user characteristics found in these communities may not be applicable to communities under other topics. Second, ``flair'' may not be accurate to reflect team affiliation for some users. For example, users may just choose the first team they watched as the flair. Third, we only select the relatively active users (having at least 5 activities) and the users who identify their team flair for our analysis. Therefore, we are not able to observe implicit team affiliation such as participating in the home team related discussion but not using flairs, and ``indirect'' active users such as the ones who only browse the content but not comment. Moreover, our dataset is limited due to the assumption of defining the "home team" of a bandwagon fan as the team they originally supported before bandwagoning, without considering the previous fan-ship behavior of the user. This definition may not accurately reflect the true nature of the fan's loyalty. A future work can be studying the methods to identify implicit team affiliation and bandwagon behavior. %, so that the target user base can be increased. 
Another direction is to understand the question ``why users choose to bandwagon to another team?". Possible reasons include finding a team to support after one's home team is eliminated, losing interest in home team, or supporting ``enemy's enemy''~\cite{bandwagonapp2019}. Answering this question will help provide a fundamental explanation to the bandwagon behavior, which could potentially help design better communities. In addition, though we have conducted sentiment analysis of bandwagon user language with the current dataset, we didn't find any significant results to differentiate different bandwagon user types due to the limitations of the dataset scope. This is another direction worth exploring as future work: the sentiments of bandwagon user language can be further analyzed and described in other application domains such as more generic sports communities, politics, etc.

%We compared different bandwagon fans in the respective seasons. Another direction is to characterize bandwagon behavior in a longitudinal fashion. Although we looked into longer term loyalty status, it only stopped at the very beginning of how they switch the flair. A longitudinal study on users' full ``lifecycle'' in sports community could help understand why/how a user become a loyal/unloyal fan, thus helping user profiling and community management. Another direction is to understand the reasons behind bandwagoning. Possible reasons include finding a team to support after one's home team is eliminated, losing interest in home team, or supporting ``enemy's enemy''~\cite{bandwagonapp2019}. Answering this question will help provide a fundamental explanation to the bandwagon behavior, which could potentially help design better communities. 

\bibliographystyle{IEEEtran}
\bibliography{reference}
\end{document}